\documentclass[aps,prl,showpacs,amsmath,amssymb,amsfonts,superscriptaddress,twocolumn,lengthcheck]{revtex4-1}
\usepackage{graphicx}
\usepackage{verbatim}
\usepackage{dcolumn}% Align table columns on decimal point
\usepackage{bm}% bold math
\usepackage{color}
\usepackage[colorlinks=true,citecolor=blue,linkcolor=blue,urlcolor=blue]{hyperref}
\usepackage{url}
\usepackage{float}

\newcommand{\bra}[1]{\left\langle #1\right|}
\newcommand{\ket}[1]{\left|#1\right\rangle}

\newcommand{\tr}[1]{\mathrm{tr}\left\{#1\right\}}

\newcommand{\td}{\mathrm{d}}

\newcommand{\e}[1]{\exp{\left(#1\right)}}

\newcommand{\bla}{bla\\bla\\bla\\bla\\bla}

\newcommand{\PRA}{Phys. Rev. A }

\newcommand{\mc}[1]{\mathcal{#1}}

\newcommand{\mrm}[1]{\mathrm{#1}}

\begin{document}
\title{Environment assisted speed-up of the field evolution in cavity QED}
%\title{Environment assisted speed up of the quantum state evolution of the field in a Cavity QED system.}
\author{A. D. Cimmarusti}
\affiliation{Joint Quantum Institute, Department of Physics, University of Maryland and National Institute of Standards and Technology, College Park, MD 20742, USA.}%
\author{ Z. Yan}
\affiliation{Joint Quantum Institute, Department of Physics, University of Maryland and National Institute of Standards and Technology, College Park, MD 20742, USA.}%
\affiliation{State Key Laboratory of Quantum Optics and Quantum Optics Devices, Institute of Opto-Electronics, Shanxi University, Taiyuan 030006, China.}%
\author{ B. D. Patterson}
\affiliation{Joint Quantum Institute, Department of Physics, University of Maryland and National Institute of Standards and Technology, College Park, MD 20742, USA.}%
\author{L. P. Corcos}
\affiliation{Joint Quantum Institute, Department of Physics, University of Maryland and National Institute of Standards and Technology, College Park, MD 20742, USA.}%
\author{L. A. Orozco}
\affiliation{Joint Quantum Institute, Department of Physics, University of Maryland and National Institute of Standards and Technology, College Park, MD 20742, USA.}%
\email{lorozco@umd.edu}
\author{S. Deffner}
\affiliation{Theoretical Division and Center for Nonlinear Studies, Los Alamos National Laboratory, Los Alamos, NM 87545, USA.}
\email{sdeffner@lanl.gov}

\begin{abstract}
We measure the quantum speed of the state evolution of the field in a weakly-driven optical cavity QED system. To this end,  the mode of the electromagnetic field is considered as a quantum system of interest with a preferential coupling to a tunable environment: the atoms. By controlling the environment, i.e., changing the number of atoms coupled to the optical cavity mode, an environment assisted speed-up is realized: the quantum speed of the state re-population in the optical cavity increases with the coupling strength between the optical cavity mode and this non-Markovian environment (the number of atoms).
\end{abstract}

\maketitle

%\section{Introduction}
%\label{sec:intro_speedup}
Identifying time-optimized processes is a ubiquitous goal in virtually all areas of quantum physics, such as quantum communication \cite{Bekenstein1981}, quantum computation \cite{lloyd00}, quantum thermodynamics \cite{Deffner2010}, quantum control and feedback \cite{caneva09}, and quantum metrology \cite{Giovannetti2011}. To this end, the notion of a quantum speed limit (QSL) has proven to be useful and important. The QSL determines the theoretical upper bound on the speed of evolution of a quantum system. It can be understood as a generalization of the Heisenberg uncertainty relation for energy and time. It has been derived for isolated, uncontrolled systems \cite{mandelstam45,margolus98,Giovannetti2004}, time-dependent Hamiltonians \cite{Barnes2013a,Poggi2013,Hegerfeldt2013,Andersson2014,Deffner2013a}, and more recently for more general open system dynamics \cite{deffner13,delcampo13,taddei13,deffner14,Zhang2014,Mukherjee2013,Xu2014,Xu2014a}. Although fundamental in nature, practical consequences or even experimental applications of the QSL are still lacking. Nevertheless,  achieving the maximal quantum speed is of high practical relevance, especially in the development of quantum information processing devices. 

On the theoretical side, a  recent study \cite{deffner13} hinted at the possibility of observing \emph{speed-ups} of the quantum evolution if an open quantum system is subject to environmental changes. Ref.~\cite{deffner13} analyzes the dynamics of the damped Jaynes-Cummings model, which describes many cavity QED systems. These systems in both the intermediate and strong coupling regime can exhibit environment-assisted evolution~\cite{madsen11} -- such as non-exponential decay.

This letter reports an experimental realization of the theoretically proposed \textit{environment assisted speed-up} \cite{deffner13}. To this end, we look at the system in an unusual way -- as just consisting of the cavity field. This allows us to treat the atomic number that generates the atomic polarization (the off diagonal elements of the atomic master equation) as a tunable environment with a range of coupling constants. We demonstrate that increasing the interaction of the optical cavity field with the environment  by tuning the number of atoms, modifies the time dependent non-classical intensity correlation function,  enhancing -- \emph{speeds-up}-- the rate of evolution of the cavity field  in a range with no clear oscillations present.

Our cavity QED system operates in the intermediate coupling regime, where the cavity-atom parameters are of the same order: $\left(g,\kappa,\gamma \right) / 2 \pi = \left(3.2,4.5,6.0\right)$ MHz. Here $g$ denotes the electric dipole interaction strength of an atom maximally coupled to the cavity mode (also known as the single photon Rabi frequency), and $\kappa$ and $\gamma$ are the decay rates of the cavity and the atom, respectively. For weak driving and $N$  atoms in the cavity \cite{carmichael91}, the effective dipole coupling between cavity and atomic environment scales as   $g\sqrt{N} \approx \Omega_{\rm VR}$
the vacuum Rabi Oscillation. The interaction of cavity and atoms is stronger for larger $N$. In contrast to conventional studies of cavity QED we make full use of this observation and its tunability by using a slow atomic beam:  The atomic beam, i.e., a collection of $N$ two-level atoms randomly positioned in the cavity mode,  with a range of position dependent coupling constants $g(r,\theta,z)$, quantified by $N_{\rm eff}$ atoms \cite{rempe91}, can be understood as a \textit{controllable, tailored environment} for the cavity mode, which we consider as the quantum system (of interest). Our atomic beam can produce a wide range of effective number of atoms in the cavity ($N_{\rm eff} = 0.1 \rightarrow 30$), which allows us to investigate the speed of state evolution  as we change the number of atoms in a controllable way.

Intuitively, one expects to observe a faster state evolution when more atoms are in the cavity;  the more subtle effect is that there is a non-linear dependence on the number of atoms. Conditional measurements of the photons leaving the cavity ($g^{(2)}(\tau)$, second order intensity correlation function) are ideal to study these environmental effects. In our experiment we measure $g^{(2)}(\tau)$ as we change the environment, the number of atoms. More concretely, we investigate the initial antibunching dynamics as the state returns to a steady state \emph{before} vacuum Rabi oscillations are present  \cite{rempe91,mielke98,foster00pra}. 

Our measurements clearly show the theoretically predicted environment-assisted speed-up with its non-linear dependence on the number of atoms. This study not only verifies fundamental predictions, but also opens new avenues %in understanding how 
to control the quantum-speed-limited dynamics in any cavity QED system. 

\paragraph{Theoretical predictions.} 

%To gain physical insight and build intuition 
We start by discussing a simple, phenomenological model for our system. To this end, we consider the optical cavity to be weakly driven by an external field of strength proportional to $\varepsilon$, which generates a field coupled to $N$ two-level atoms through the Cooperativity $C=NC_1,~C_1=g^2/\kappa\gamma$. This situation is described by the master equation \cite{carmichael91}
\begin{equation}
\label{eq:master}
\begin{split}
&\dot{\rho}(t)=\varepsilon\left[a^\dagger-a,\rho(t)\right]+ g\left[a^\dagger J_- -a J_+,\rho(t) \right]\\
&\hspace{.5em}+ \kappa \left(2 a\,\rho(t)\, a^\dagger-a^\dagger a \,\rho(t) -\rho(t)\, a^\dagger a\right)\\
&\hspace{.5em}+\gamma/2\sum\limits_{j=1}^N \left(2 \sigma^j_-\,\rho(t)\, \sigma^j_+ - \sigma^j_+ \sigma^j_-\, \rho(t) -\rho(t)\, \sigma^j_+ \sigma^j_-\right).
\end{split}
\end{equation}
Here $a^\dagger$ and $a$ describe the  creation and annihilation operators for the cavity mode, $\sigma^j_+$ and $\sigma^j_-$ correspond to the Pauli spin matrices for the $j$th atom, while
 $J_\pm=\sum_{j=1}^N \sigma^j_\pm$ are collective excitations. In the limit of weak driving, $\varepsilon/\kappa \ll 1$, a perturbative treatment is possible. Then, a solution of Eq.~\eqref{eq:master} can be approximated by $\rho(t)\simeq\ket{\psi(t)}\bra{\psi(t)}$ with
\begin{equation}
\label{eq:state}
 \ket{\psi(t)}\simeq\ket{00}+A_1(t)\,\ket{10}+A_2(t)\, \ket{01}+\mc{O}(\varepsilon^2/\kappa^2)
\end{equation}
where $\ket{nm}$ state with $n$ photons in cavity and $m$ excited atoms. The dynamics of the amplitudes $A_1(t)$ and $A_2(t)$ are described by \cite{carmichael91}
\begin{equation}
\label{eq:simple}
\begin{split}
\dot{A_1}(t)&=-\kappa \,A_1(t)+g\sqrt{N} \, A_2(t)+\varepsilon\\
\dot{A_2}(t)&=-\gamma/2 \,A_2(t)-g\sqrt{N} \,A_1(t)\,.
\end{split}
\end{equation}
Adiabatically eliminating the cavity $\kappa \ll \gamma, g\sqrt{N}$ one gets the Purcell enhancement factor that changes the exponential decay rate of the atoms from $\gamma$ to $\gamma(1+2C)$  which depends linearly on $N$. A similar result for the cavity decay rate is found by adiabatically eliminating the atoms $\gamma \ll \kappa, g\sqrt{N}$ where now the decay rate of the cavity changes from $\kappa$ to $\kappa(1+2C)$ again linearly dependent on $N$.
Here we want to stress a different regime, where we keep both entities in the evolution and so  we immediately observe that the cavity field couples to $N$ two-level atoms through a  dipole transition with rate $g \sqrt{N}$   -- an effect not indicated by oscillations. Thus, by tuning the number of atoms $N$ the interaction strength of the cavity mode and the atoms can be varied. In the following, we focus exclusively on the escaping field of the system as we vary the number of atoms. Thus, we can monitor the changes in its dynamic behavior as we modify this ``variable reservoir'' -- showing the potential of engineering dynamical behavior by manipulating the reservoir.

Reference~\cite{carmichael91} also derived the solution of Eq.~\eqref{eq:simple} and a closed expression for the second order correlation function $g^{(2)}(\tau)$ applicable for $N$ atoms, which reads
\begin{equation}
\label{eq:g2_final}
\begin{split}
  &g^{(2)}(\tau)  =  \bigg\{1 + \frac{\Delta \alpha}{\alpha} \e{-\frac{\tau}{2}\,\left(\kappa + \frac{\gamma}{2}\right)}  \\
                        & \quad\times \left[ \cosh{\left(\Omega_\mrm{VR}\,\tau\right)} + \frac{1}{2} \left(\kappa + \frac{\gamma}{2}\right)\, \frac{\sinh{\left(\Omega_\mrm{VR}\, \tau\right)}}{\Omega_\mrm{VR}} \right]  \bigg\}^2 
\end{split}
\end{equation}
where the parameters are defined by, $\Delta \alpha/\alpha = - 2 C'_1 (2 C/(1 + 2 C - 2 C'_1))$, and $C'_1=C_1/(1+\gamma/2\kappa)$. Furthermore, the vacuum Rabi frequency is,  $\Omega_\mrm{VR} = \sqrt{ \left(( \kappa - \gamma/2 \right)/2)^2 - g^2 N}$.
\begin{figure}
\includegraphics[width=.48\textwidth]{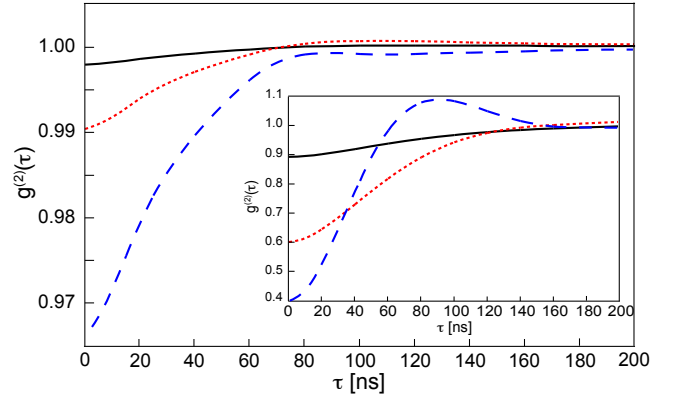}
\caption{\label{fig:g2} (color online) Calculated correlation function $g^{(2)}(\tau)$ \eqref{eq:g2_final} accounting for experimental effects for $\left(g,\kappa,\gamma \right) / 2 \pi = \left(3.2,4.5,6.0\right)$ MHz with $\Omega_{VR}/2\pi=1.1$ MHz (black, continuous line), 2.8 MHz (red dotted line), and 5.2 (blue, dashed line). The inset  corresponds to a model with maximally coupled atoms and optimal experimental conditions.}
\end{figure}

This  master equation model \eqref{eq:master} assumes static, maximally coupled atoms to the cavity field with zero cavity and atomic detuning. A refined model simulates fluctuations of the atomic beam number with a Poissonian weight and randomly generates the positions of the atoms in the cavity mode (which is radially Gaussian with a longitudinal standing wave)~\cite{thompson92, carmichael99}. This leads to variable dipole coupling strengths and effectively decreases the overall coupling. The atoms present then a range of couplings and behave as a reservoir depending on where they are located on the mode. The calculation also includes experimental effects from cavity and atomic detuning of $g^2(\tau)$ is plotted in Fig.~\ref{fig:g2}  for the corresponding $\Omega_{VR}$~\cite{brecha99,foster00pra}. We observe that while the Rabi oscillations are strongly suppressed the rate of refilling characterized by the slope at the inflection point, becomes a characteristic for the state dynamics returning to the steady state. The atoms, with their different couplings constants to the cavity mode behave as an inhomogeneously broadened reservoir, with a variety of coupling constants that nevertheless preserve the non-classical anti-bunched nature of the field. All the oscillations start anti-bunched but their oscillations are different and averaging them produces dephasing that results in a simple decay rate that will preserve the non-linear $\sqrt{N}$ dependence.  The inhomogenous reservoir is crucial for this discussion. The inset in Fig~\ref{fig:g2} shows the result of $N$ maximally coupled atoms and optimal experimental conditions.

Before we continue the discussion with the experimental results, note that the  speed-up proposed here is physically equivalent to the behavior of the QSL predicted in Ref.~\cite{deffner13}. However, here we are interested in the dynamics of the field, whereas Ref.~\cite{deffner13} focused on the atoms. 

\paragraph{Experimental set-up and results.}

\begin{figure}[h]
\begin{center}
\includegraphics[width=0.9\linewidth]{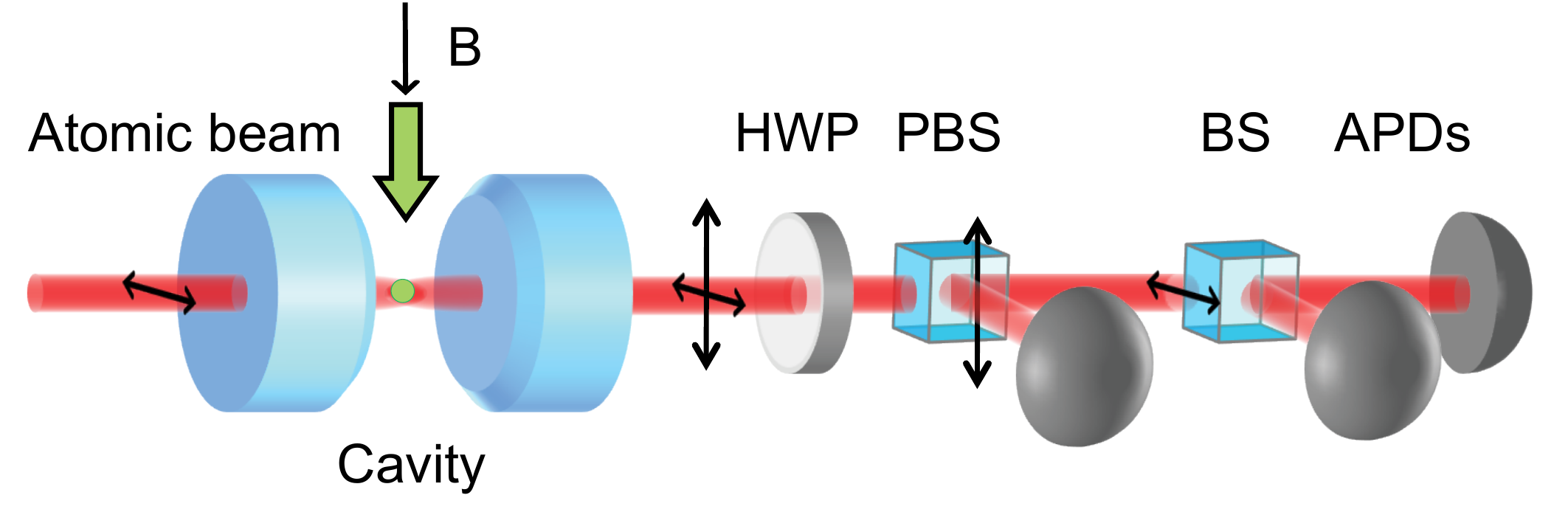}
\caption{\label{fig:speedup_exp} (color online) Simplified diagram of the apparatus for measuring autocorrelations with an atomic beam in the presence of a magnetic field B. HWP: half wave plate; PBS: polarizing beam splitter; BS: beam splitter; APD: avalanche photo-diode.}
\end{center}
\end{figure}

Figure~\ref{fig:speedup_exp} shows the general layout of the physical system, which is our new-generation optical cavity QED apparatus based on the previous one described in Ref.~\cite{norris09a} operating on the D$_2$ line of Rb. The new system has $n_{\rm sat} \equiv \gamma^2 /3g^2 =1.2$ and $C_1 \equiv g^2/\kappa\gamma = 0.38$. An imbalanced magneto optical trap (MOT) produces a slow atomic beam of $^{85}$Rb atoms that travels down and couples to the TEM$_{00}$ mode of our 0.8 mm long optical cavity. The magnetic field points along the $-z$ axis and defines our quantization axis. The MOT coils and an arrangement of four permanent magnets generate this field of about $\vec{B} = -7.2$ G $\hat{z}$.

The downward pointing  laser beam from the MOT provides optical pumping into the $\lvert F=3, m_F = 3 \rangle$ state. We verify this by measuring the  absorption as a function of frequency, keeping the cavity resonant to the laser frequency.
 We drive the system with $H$-polarized light (perpendicular to the B magnetic field). In the frame of the atoms, it appears as a combination of  two circularly polarized fields, while only one is capable of driving with the stretched transition.

Experimentally, we increase the number of atoms by raising the current of the Rb dispenser, which increases the number of atoms in the MOT. 
A locked 820 nm laser serves to keep the cavity on resonance with the atoms and the excitation laser at 780 nm.

Polarization elements and mode-matching optics prepare the driving laser before it enters the cavity. A lens at the output collimates the beam and a half-wave plate (HWP) aligns the polarization to a calcite PBS which separates the H (driven) and V (undriven) modes. The H mode passes through a second beam splitter which divides the light between two avalanche photodiodes (APD) for single or coincidence measurements. A series of frequency and spatial filters remove background light. The APD electronic output pulses then go to a time-stamp unit in a computer which, after post-detection signal processing, produces the appropriate correlations. The total detection efficiency of the system is 30\%. We block part of the laser beam that could co-propagate with the atomic beam either with a dark spot or with an axicon system that produces a dark region at the cavity \cite{kulin01}.
The geometry and the detection that we use place limitations on the state purity and signal background. Nevertheless, we can identify the non-classical feature of antibunching and measure the time the system needs to relax back to its steady state. 

We measure $\Omega_{\rm VR}(N_{\rm eff}) $ in the cavity QED system by transmission spectroscopy for a given the atomic flux. We scan the drive frequency while keeping the atoms and the cavity resonant \cite{zhu90,gripp97}. The separation between the two peaks is twice $\Omega_{\rm VR}$. We observe a clear splitting on the transmission spectroscopy while on the correlation functions the corresponding time oscillations are not present \cite{foster00pra} for the range of parameters that we explore in this paper; however we do observe the oscillatory regime for higher dispenser currents in the MOT. This is related to the different kinds of averaging that take place in the transmission spectra and in the correlation functions. 

\begin{figure}[h]
  \begin{centering}
    \includegraphics[width=0.95\linewidth]{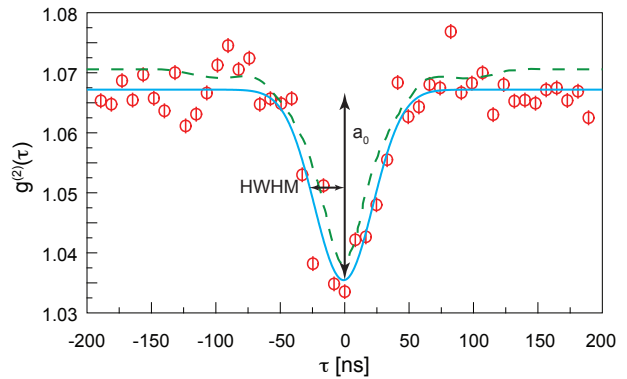}
    \caption{(color online) Measured  $g^{(2)}(\tau)$ (red circles with only statistical error bars) showing antibunching for  $\Omega_{\rm{VR}}\sqrt{N_{\rm eff} }= 8.5 \pm 0.1$ MHz with fit to an inverted Lorenzian (blue, solid line) with the amplitude $a_0$ and the HWHM indicated, as well as the prediction from the refined model (green, dashed line).}
    \label{fig:antibunch_fit}
  \end{centering}
\end{figure}

Figure~\ref{fig:antibunch_fit} shows one of our experimental traces of $g^{(2)}(\tau)$. We extract the quantum speed from the rate of change of the anti-bunching towards the steady state. $g^2(\tau)$ reaches unity only after a long time, which is a remnant of the transit time of the atoms through the cavity mode \cite{foster00pra}. We fit an inverted Loretzian  (continuous blue line), with three parameters: offset, amplitude $a_0$, and the half width, half maximum (HWHM).  
This gives a quantitative measurement of the rate of refilling though the slope by taking the ratio $a_0/{\rm HWHM}$, %{which is subject to systematic effects from residual backgrounds, and hence quantitative comparisons to models are premature.} 
 The dashed green line shows the results of the refined model (as for Fig.~\ref{fig:g2}) that includes experimental fluctuations on the cavity detunings ($\pm 2.5\kappa$), the presence of a Zeeman sub level 5 MHz away, and the atomic spatial distribution. The amplitude is scaled to take into account experimental backgrounds. Spurious electronic noise makes the fluctuations larger than purely statistical.

\begin{figure}
  \begin{centering}
  \includegraphics[width=0.95\linewidth]{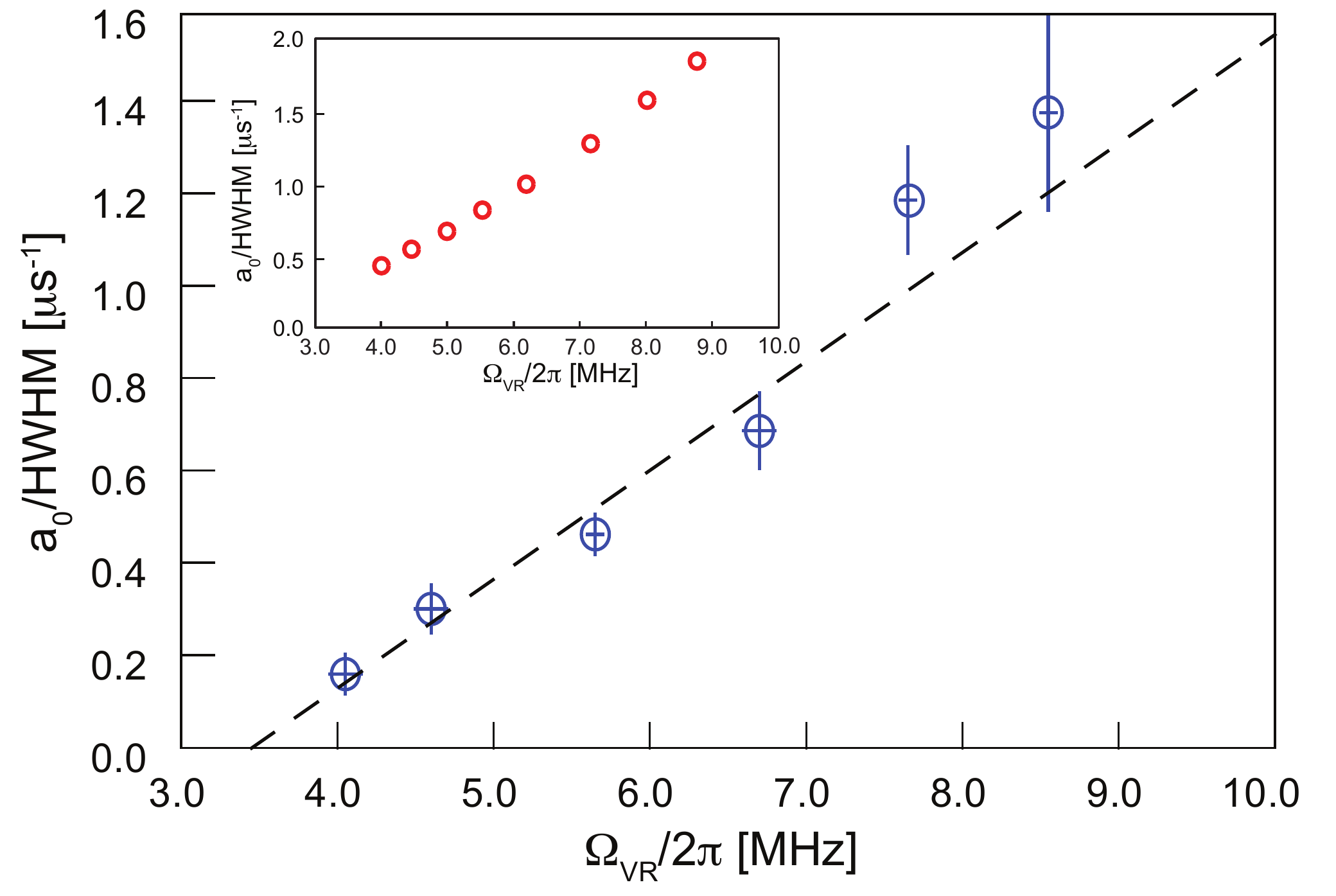}
  
\caption{(color online) Speed at the inflection point,  $a_0$/HWHM, (blue circles) as function of $\Omega_{\rm VR}$ from measured antibunching curves. Inset $a_0/{\rm HWHM}$ from the refined model (red dots) as in Figs.~\ref{fig:g2}~and~\ref{fig:antibunch_fit}.}
  \label{fig:data}
\end{centering}
\end{figure}

Figure~\ref{fig:data} shows the  growth of the rate of refilling (blue circles) as a function of $\Omega_{\rm VR}$, for which we expect a linear dependence. The dashed straight line is a fit with slope $0.24 \pm 0.03$ $\mu$s$^{-1}$/MHz, with reduced $\chi^2 = 1.67$. The inset (red dots) shows the result of a simulation with slope  $ 0.29 \pm 0.02$ $\mu$s$^{-1}$/MHz, of the refined model. 

Our results clearly indicate a trend of enhanced evolution speed as a function $\Omega_{\rm VR} \approx g\sqrt{N_{\rm eff}}$. This is in agreement with theoretical predictions~\cite{deffner13} that observe a speed-up as the coupling increases in a cavity QED system and in contrast to the dependence of $N$ in the adiabatically eliminated cases stated above, that do not show in the non-classical correlation function.  Additional measurements show a similar linear behavior of the slope at the inflection point when we have a 
bunching peak. The speed increases linearly as a function of $\Omega_{\rm VR}$. 

\paragraph{Non-Markovian speed-up.}
In the example studied in Ref.~\cite{deffner13} the quantum speed-up  in an open quantum system was attributed to the non-Markovian nature of the environment. The obvious question remains whether the here reported speed-up is also a non-Markovian effect.

Generally, quantifying non-Markovianity is complicated \cite{brune96,breuer07book,madsen11}. However, for simple quantum systems such as the Jaynes-Cummings model the situation drastically simplifies and a measure of non-Markovianity (${\cal{N}}$) can be introduced \cite{breuer09}. It is defined as
\begin{equation}
\label{eq:nonMarkov}
\mc{N}=\max_{\rho_{1,2}(0)}{\int_{\sigma>0}\td t\, \sigma(t,\rho_{1,2}(0))}
\end{equation}
where $\rho_{1,2}(0)$ are two initial states and
\begin{equation}
\label{eq:trace}
\sigma(t,\rho_{1,2}(0))=\frac{1}{2}\frac{\td}{\td t}\tr{\left|\rho_1(t)-\rho_2(t)\right|}\,.
\end{equation}
It has been seen \cite{breuer09} that for Markovian dynamics all initial states monotonically converge towards a unique stationary state. Thus, $\sigma(t,\rho_{1,2}(0))$ \eqref{eq:trace} is strictly negative and $\mc{N}=0$ \eqref{eq:nonMarkov}. Non-Markovian dynamics are characterized by an information backflow from the environment, and the convergence of $\rho(t)$ towards the stationary state is accompanied by oscillations. Hence, $\sigma(t,\rho_{1,2}(0))$ \eqref{eq:trace} can become positive, which amounts to finite values of $\mc{N} $.

\begin{figure}[h]
\begin{center}
\includegraphics[width=0.9\linewidth]{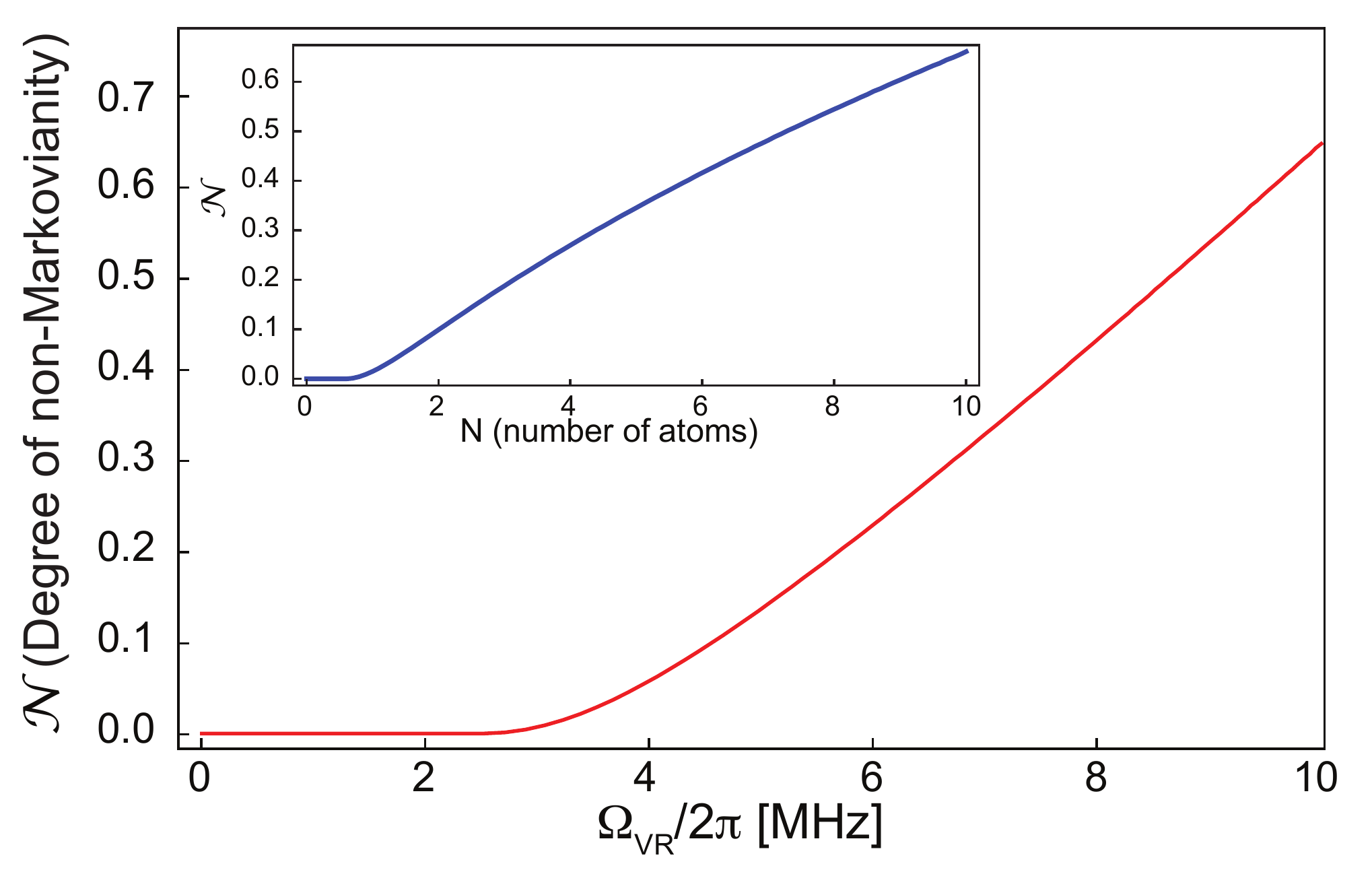}
\caption{\label{fig:Non_Mark} (color online)  Measure of non-Markovianity (red line) in the cavity QED system \eqref{eq:nonMarkov} averaged over a distribution of atoms in the cavity mode as a function of the Vacuum Rabi frequency. The inset (blue line) shows the same $\mc{N}$ for maximally coupled atoms. }
\end{center}
\end{figure}

Figure~\ref{fig:Non_Mark} shows ${\cal{N}}$ \eqref{eq:nonMarkov} for the phenomenological model \eqref{eq:simple}. The red line is the result of an average over the distribution of the atoms in the mode while the inset with the blue line  assumes maximally coupled atoms with fixed positions. The averaged line shows a linear dependence with the Vacuum Rabi frequency that is proportional to $\sqrt{N_{\rm eff}}$. We observe that the non-Markovianity increases with the number of atoms, and that there is a threshold for starting to observe non-Markovian behavior. This threshold can be understood intuitively: To have non-Markovian behavior the environment has to be correlated; to be correlated the environment has to consist at least of one atom.  This threshold also bears an important message on the nature of non-Markovianity for open quantum systems: The presence of oscillations in the system (non-zero vacuum Rabi oscillations) is not a good criterion for non-Markovian behavior. Even in the non-averaged plot (inset of Fig.~\ref{fig:Non_Mark}) the threshold at $N\simeq 1$ to observe non-Markovianity is clearly visible. On the contrary, Rabi oscillations are already present for $N\ll 1$.

Figure~\ref{fig:Non_Mark} also confirms that the reported speed-up is indeed a non-Markovian effect since the speed of evolution grows with the number of atoms, which grows with the non-Markovianity \cite{Note2}.
%\footnote{It is worth emphasizing that in our study increase of non-Markovianity is a necessary, but not sufficient condition to observe a speed-up. See also Ref.~\cite{Xu2014}}. 
Note that Fig.~\ref{fig:Non_Mark} also indicates that the non-Markovianity survives the averaging over the random distribution of atoms. 

\paragraph{Concluding remarks.}

The reported experiment %realized three major achievements: 
first
demonstrated the theoretically predicted environment assisted speed-up. Second,% by changing the common point of view 
by considering the cavity field as a system coupled to a tunable environment realized a fully accessible quantum system obeying non-Markovian dynamics. Third demonstrated a novel way of environment engineering %that marks a path 
for optimal quantum control protocols~\cite{deffner14}. 

The spatial distribution of the atoms, producing a broadband of couplings, is playing a major role in allowing us to treat them as a non-Markovian reservoir as we do see the predicted $\sqrt{N}$ behavior.
% and not the trivial result of the adiabatic elimination of the atoms that produces a Markovian result. 

These findings could prove useful to realize strong feedback applied, e.g., in capture and release of a conditional state of cavity QED~\cite{smith02} and for a spontaneously created coherence~\cite{cimmarusti13}. %The use of conditional measurements, quantum intensity correlations, allows us to follow the evolution of the conditional state in a unique manner.
In future work we will explore the quantum speed in optical cavity QED for large $N$ to investigate any possible functional changes on the speed as suggested by Taddei {\it et al.} \cite{taddei13}. A measurement of the cross correlation between the field and the atomic polarization can give a complementary picture, and we are developing ways to do it.
We finally note that the  effect reported here is generic in the sense that it is not a peculiarity of two-level systems and we expect to go beyond it in forthcoming experiments. 
%This point will be further elucidated in a forth-coming experiment in which we will go beyond the two-level atom and use the orthogonal mode as we have done for ground state coherences \cite{norris10}.

\acknowledgments
 We would like to thank H. J. Carmichael for helpful discussions and P. Dussarrat for help with the apparatus.
The work at UMD is supported by grant 12 NSF of the USA (Grant No. 1307416). Z. Y. acknowledges support from the Natural Science Foundation of China (Grant  11304190). S. D. acknowledges financial support by the U.S. Department of Energy through a LANL Director's Funded Fellowship.


\begin{thebibliography}{10}
\makeatletter
\providecommand \@ifxundefined [1]{%
 \@ifx{#1\undefined}
}%
\providecommand \@ifnum [1]{%
 \ifnum #1\expandafter \@firstoftwo
 \else \expandafter \@secondoftwo
 \fi
}%
\providecommand \@ifx [1]{%
 \ifx #1\expandafter \@firstoftwo
 \else \expandafter \@secondoftwo
 \fi
}%
\providecommand \natexlab [1]{#1}%
\providecommand \enquote  [1]{``#1''}%
\providecommand \bibnamefont  [1]{#1}%
\providecommand \bibfnamefont [1]{#1}%
\providecommand \citenamefont [1]{#1}%
\providecommand \href@noop [0]{\@secondoftwo}%
\providecommand \href [0]{\begingroup \@sanitize@url \@href}%
\providecommand \@href[1]{\@@startlink{#1}\@@href}%
\providecommand \@@href[1]{\endgroup#1\@@endlink}%
\providecommand \@sanitize@url [0]{\catcode `\\12\catcode `\$12\catcode
  `\&12\catcode `\#12\catcode `\^12\catcode `\_12\catcode `\%12\relax}%
\providecommand \@@startlink[1]{}%
\providecommand \@@endlink[0]{}%
\providecommand \url  [0]{\begingroup\@sanitize@url \@url }%
\providecommand \@url [1]{\endgroup\@href {#1}{\urlprefix }}%
\providecommand \urlprefix  [0]{URL }%
\providecommand \Eprint [0]{\href }%
\providecommand \doibase [0]{http://dx.doi.org/}%
\providecommand \selectlanguage [0]{\@gobble}%
\providecommand \bibinfo  [0]{\@secondoftwo}%
\providecommand \bibfield  [0]{\@secondoftwo}%
\providecommand \translation [1]{[#1]}%
\providecommand \BibitemOpen [0]{}%
\providecommand \bibitemStop [0]{}%
\providecommand \bibitemNoStop [0]{.\EOS\space}%
\providecommand \EOS [0]{\spacefactor3000\relax}%
\providecommand \BibitemShut  [1]{\csname bibitem#1\endcsname}%
\let\auto@bib@innerbib\@empty
%</preamble>
\bibitem [{\citenamefont {Bekenstein}(1981)}]{Bekenstein1981}%
  \BibitemOpen
  \bibfield  {author} {\bibinfo {author} {\bibfnamefont {J.~D.}\ \bibnamefont
  {Bekenstein}},\ }\href {\doibase 10.1103/PhysRevLett.46.623} {\bibfield
  {journal} {\bibinfo  {journal} {Phys. Rev. Lett.}\ }\textbf {\bibinfo
  {volume} {46}},\ \bibinfo {pages} {623} (\bibinfo {year} {1981})}\BibitemShut
  {NoStop}%
\bibitem [{\citenamefont {Lloyd}(2000)}]{lloyd00}%
  \BibitemOpen
  \bibfield  {author} {\bibinfo {author} {\bibfnamefont {S.}~\bibnamefont
  {Lloyd}},\ }\href {\doibase 10.1038/35023282} {\bibfield  {journal} {\bibinfo
   {journal} {Nature}\ }\textbf {\bibinfo {volume} {406}},\ \bibinfo {pages}
  {1047} (\bibinfo {year} {2000})}\BibitemShut {NoStop}%
\bibitem [{\citenamefont {Deffner}\ and\ \citenamefont
  {Lutz}(2010)}]{Deffner2010}%
  \BibitemOpen
  \bibfield  {author} {\bibinfo {author} {\bibfnamefont {S.}~\bibnamefont
  {Deffner}}\ and\ \bibinfo {author} {\bibfnamefont {E.}~\bibnamefont {Lutz}},\
  }\href {\doibase 10.1103/PhysRevLett.105.170402} {\bibfield  {journal}
  {\bibinfo  {journal} {Phys. Rev. Lett.}\ }\textbf {\bibinfo {volume} {105}},\
  \bibinfo {pages} {170402} (\bibinfo {year} {2010})}\BibitemShut {NoStop}%
\bibitem [{\citenamefont {Caneva}\ \emph {et~al.}(2009)\citenamefont {Caneva},
  \citenamefont {Murphy}, \citenamefont {Calarco}, \citenamefont {Fazio},
  \citenamefont {Montangero}, \citenamefont {Giovannetti},\ and\ \citenamefont
  {Santoro}}]{caneva09}%
  \BibitemOpen
  \bibfield  {author} {\bibinfo {author} {\bibfnamefont {T.}~\bibnamefont
  {Caneva}}, \bibinfo {author} {\bibfnamefont {M.}~\bibnamefont {Murphy}},
  \bibinfo {author} {\bibfnamefont {T.}~\bibnamefont {Calarco}}, \bibinfo
  {author} {\bibfnamefont {R.}~\bibnamefont {Fazio}}, \bibinfo {author}
  {\bibfnamefont {S.}~\bibnamefont {Montangero}}, \bibinfo {author}
  {\bibfnamefont {V.}~\bibnamefont {Giovannetti}}, \ and\ \bibinfo {author}
  {\bibfnamefont {G.~E.}\ \bibnamefont {Santoro}},\ }\href {\doibase
  10.1103/PhysRevLett.103.240501} {\bibfield  {journal} {\bibinfo  {journal}
  {Phys. Rev. Lett.}\ }\textbf {\bibinfo {volume} {103}},\ \bibinfo {pages}
  {240501} (\bibinfo {year} {2009})}\BibitemShut {NoStop}%
\bibitem [{\citenamefont {Giovannetti}\ \emph {et~al.}(2011)\citenamefont
  {Giovannetti}, \citenamefont {Lloyd},\ and\ \citenamefont
  {Maccone}}]{Giovannetti2011}%
  \BibitemOpen
  \bibfield  {author} {\bibinfo {author} {\bibfnamefont {V.}~\bibnamefont
  {Giovannetti}}, \bibinfo {author} {\bibfnamefont {S.}~\bibnamefont {Lloyd}},
  \ and\ \bibinfo {author} {\bibfnamefont {L.}~\bibnamefont {Maccone}},\ }\href
  {\doibase 10.1038/nphoton.2011.35} {\bibfield  {journal} {\bibinfo  {journal}
  {Nat. Photonics}\ }\textbf {\bibinfo {volume} {5}},\ \bibinfo {pages} {222}
  (\bibinfo {year} {2011})}\BibitemShut {NoStop}%
\bibitem [{\citenamefont {Mandelstam}\ and\ \citenamefont
  {Tamm}(1945)}]{mandelstam45}%
  \BibitemOpen
  \bibfield  {author} {\bibinfo {author} {\bibfnamefont {L.}~\bibnamefont
  {Mandelstam}}\ and\ \bibinfo {author} {\bibfnamefont {I.}~\bibnamefont
  {Tamm}},\ }\href@noop {} {\bibfield  {journal} {\bibinfo  {journal} {J.
  Phys.}\ }\textbf {\bibinfo {volume} {9}},\ \bibinfo {pages} {249} (\bibinfo
  {year} {1945})}\BibitemShut {NoStop}%
\bibitem [{\citenamefont {Margolus}\ and\ \citenamefont
  {Levitin}(1998)}]{margolus98}%
  \BibitemOpen
  \bibfield  {author} {\bibinfo {author} {\bibfnamefont {N.}~\bibnamefont
  {Margolus}}\ and\ \bibinfo {author} {\bibfnamefont {L.~B.}\ \bibnamefont
  {Levitin}},\ }\href {\doibase 10.1016/S0167-2789(98)00054-2} {\bibfield
  {journal} {\bibinfo  {journal} {Phys. D}\ }\textbf {\bibinfo {volume}
  {120}},\ \bibinfo {pages} {188} (\bibinfo {year} {1998})}\BibitemShut
  {NoStop}%
\bibitem [{\citenamefont {{Giovannetti}}\ \emph {et~al.}(2004)\citenamefont
  {{Giovannetti}}, \citenamefont {{Lloyd}},\ and\ \citenamefont
  {{Maccone}}}]{Giovannetti2004}%
  \BibitemOpen
  \bibfield  {author} {\bibinfo {author} {\bibfnamefont {V.}~\bibnamefont
  {{Giovannetti}}}, \bibinfo {author} {\bibfnamefont {S.}~\bibnamefont
  {{Lloyd}}}, \ and\ \bibinfo {author} {\bibfnamefont {L.}~\bibnamefont
  {{Maccone}}},\ }\href {\doibase 10.1088/1464-4266/6/8/028} {\bibfield
  {journal} {\bibinfo  {journal} {J. Opt. B}\ }\textbf {\bibinfo {volume}
  {6}},\ \bibinfo {pages} {807} (\bibinfo {year} {2004})}\BibitemShut {NoStop}%
\bibitem [{\citenamefont {Barnes}(2013)}]{Barnes2013a}%
  \BibitemOpen
  \bibfield  {author} {\bibinfo {author} {\bibfnamefont {E.}~\bibnamefont
  {Barnes}},\ }\href {\doibase 10.1103/PhysRevA.88.013818} {\bibfield
  {journal} {\bibinfo  {journal} {Phys. Rev. A}\ }\textbf {\bibinfo {volume}
  {88}},\ \bibinfo {pages} {013818} (\bibinfo {year} {2013})}\BibitemShut
  {NoStop}%
\bibitem [{\citenamefont {{Poggi}}\ \emph {et~al.}(2013)\citenamefont
  {{Poggi}}, \citenamefont {{Lombardo}},\ and\ \citenamefont
  {{Wisniacki}}}]{Poggi2013}%
  \BibitemOpen
  \bibfield  {author} {\bibinfo {author} {\bibfnamefont {P.~M.}\ \bibnamefont
  {{Poggi}}}, \bibinfo {author} {\bibfnamefont {F.~C.}\ \bibnamefont
  {{Lombardo}}}, \ and\ \bibinfo {author} {\bibfnamefont {D.~A.}\ \bibnamefont
  {{Wisniacki}}},\ }\href {\doibase 10.1209/0295-5075/104/40005} {\bibfield
  {journal} {\bibinfo  {journal} {EPL (Europhysics Letters)}\ }\textbf
  {\bibinfo {volume} {104}},\ \bibinfo {eid} {40005} (\bibinfo {year}
  {2013})}\BibitemShut {NoStop}%
\bibitem [{\citenamefont {Hegerfeldt}(2013)}]{Hegerfeldt2013}%
  \BibitemOpen
  \bibfield  {author} {\bibinfo {author} {\bibfnamefont {G.~C.}\ \bibnamefont
  {Hegerfeldt}},\ }\href {\doibase 10.1103/PhysRevLett.111.260501} {\bibfield
  {journal} {\bibinfo  {journal} {Phys. Rev. Lett.}\ }\textbf {\bibinfo
  {volume} {111}},\ \bibinfo {pages} {260501} (\bibinfo {year}
  {2013})}\BibitemShut {NoStop}%
\bibitem [{\citenamefont {{Andersson}}\ and\ \citenamefont
  {{Heydari}}(2014)}]{Andersson2014}%
  \BibitemOpen
  \bibfield  {author} {\bibinfo {author} {\bibfnamefont {O.}~\bibnamefont
  {{Andersson}}}\ and\ \bibinfo {author} {\bibfnamefont {H.}~\bibnamefont
  {{Heydari}}},\ }\href {\doibase 10.1088/1751-8113/47/21/215301} {\bibfield
  {journal} {\bibinfo  {journal} {J. Phys. A}\ }\textbf {\bibinfo {volume}
  {47}},\ \bibinfo {eid} {215301} (\bibinfo {year} {2014})}\BibitemShut
  {NoStop}%
\bibitem [{\citenamefont {{Deffner}}\ and\ \citenamefont
  {{Lutz}}(2013)}]{Deffner2013a}%
  \BibitemOpen
  \bibfield  {author} {\bibinfo {author} {\bibfnamefont {S.}~\bibnamefont
  {{Deffner}}}\ and\ \bibinfo {author} {\bibfnamefont {E.}~\bibnamefont
  {{Lutz}}},\ }\href {\doibase 10.1088/1751-8113/46/33/335302} {\bibfield
  {journal} {\bibinfo  {journal} {J. Phys. A}\ }\textbf {\bibinfo {volume}
  {46}},\ \bibinfo {eid} {335302} (\bibinfo {year} {2013})}\BibitemShut
  {NoStop}%
\bibitem [{\citenamefont {Deffner}\ and\ \citenamefont
  {Lutz}(2013)}]{deffner13}%
  \BibitemOpen
  \bibfield  {author} {\bibinfo {author} {\bibfnamefont {S.}~\bibnamefont
  {Deffner}}\ and\ \bibinfo {author} {\bibfnamefont {E.}~\bibnamefont {Lutz}},\
  }\href {\doibase 10.1103/PhysRevLett.111.010402} {\bibfield  {journal}
  {\bibinfo  {journal} {Phys. Rev. Lett.}\ }\textbf {\bibinfo {volume} {111}},\
  \bibinfo {pages} {010402} (\bibinfo {year} {2013})}\BibitemShut {NoStop}%
\bibitem [{\citenamefont {del Campo}\ \emph {et~al.}(2013)\citenamefont {del
  Campo}, \citenamefont {Egusquiza}, \citenamefont {Plenio},\ and\
  \citenamefont {Huelga}}]{delcampo13}%
  \BibitemOpen
  \bibfield  {author} {\bibinfo {author} {\bibfnamefont {A.}~\bibnamefont {del
  Campo}}, \bibinfo {author} {\bibfnamefont {I.~L.}\ \bibnamefont {Egusquiza}},
  \bibinfo {author} {\bibfnamefont {M.~B.}\ \bibnamefont {Plenio}}, \ and\
  \bibinfo {author} {\bibfnamefont {S.~F.}\ \bibnamefont {Huelga}},\ }\href
  {\doibase 10.1103/PhysRevLett.110.050403} {\bibfield  {journal} {\bibinfo
  {journal} {Phys. Rev. Lett.}\ }\textbf {\bibinfo {volume} {110}},\ \bibinfo
  {pages} {050403} (\bibinfo {year} {2013})}\BibitemShut {NoStop}%
\bibitem [{\citenamefont {Taddei}\ \emph {et~al.}(2013)\citenamefont {Taddei},
  \citenamefont {Escher}, \citenamefont {Davidovich},\ and\ \citenamefont
  {de~Matos~Filho}}]{taddei13}%
  \BibitemOpen
  \bibfield  {author} {\bibinfo {author} {\bibfnamefont {M.~M.}\ \bibnamefont
  {Taddei}}, \bibinfo {author} {\bibfnamefont {B.~M.}\ \bibnamefont {Escher}},
  \bibinfo {author} {\bibfnamefont {L.}~\bibnamefont {Davidovich}}, \ and\
  \bibinfo {author} {\bibfnamefont {R.~L.}\ \bibnamefont {de~Matos~Filho}},\
  }\href {\doibase 10.1103/PhysRevLett.110.050402} {\bibfield  {journal}
  {\bibinfo  {journal} {Phys. Rev. Lett.}\ }\textbf {\bibinfo {volume} {110}},\
  \bibinfo {pages} {050402} (\bibinfo {year} {2013})}\BibitemShut {NoStop}%
\bibitem [{\citenamefont {{Deffner}}(2014)}]{deffner14}%
  \BibitemOpen
  \bibfield  {author} {\bibinfo {author} {\bibfnamefont {S.}~\bibnamefont
  {{Deffner}}},\ }\href {\doibase 10.1088/0953-4075/47/14/145502} {\bibfield
  {journal} {\bibinfo  {journal} {J. Phys. B}\ }\textbf {\bibinfo {volume}
  {47}},\ \bibinfo {eid} {145502} (\bibinfo {year} {2014})}\BibitemShut
  {NoStop}%
\bibitem [{\citenamefont {{Zhang}}\ \emph {et~al.}(2014)\citenamefont
  {{Zhang}}, \citenamefont {{Han}}, \citenamefont {{Xia}}, \citenamefont
  {{Cao}},\ and\ \citenamefont {{Fan}}}]{Zhang2014}%
  \BibitemOpen
  \bibfield  {author} {\bibinfo {author} {\bibfnamefont {Y.-J.}\ \bibnamefont
  {{Zhang}}}, \bibinfo {author} {\bibfnamefont {W.}~\bibnamefont {{Han}}},
  \bibinfo {author} {\bibfnamefont {Y.-J.}\ \bibnamefont {{Xia}}}, \bibinfo
  {author} {\bibfnamefont {J.-P.}\ \bibnamefont {{Cao}}}, \ and\ \bibinfo
  {author} {\bibfnamefont {H.}~\bibnamefont {{Fan}}},\ }\href {\doibase
  10.1038/srep04890} {\bibfield  {journal} {\bibinfo  {journal} {Sci. Rep.}\
  }\textbf {\bibinfo {volume} {4}},\ \bibinfo {eid} {4890} (\bibinfo {year}
  {2014})}\BibitemShut {NoStop}%
\bibitem [{\citenamefont {{Mukherjee}}\ \emph {et~al.}(2013)\citenamefont
  {{Mukherjee}}, \citenamefont {{Carlini}}, \citenamefont {{Mari}},
  \citenamefont {{Caneva}}, \citenamefont {{Montangero}}, \citenamefont
  {{Calarco}}, \citenamefont {{Fazio}},\ and\ \citenamefont
  {{Giovannetti}}}]{Mukherjee2013}%
  \BibitemOpen
  \bibfield  {author} {\bibinfo {author} {\bibfnamefont {V.}~\bibnamefont
  {{Mukherjee}}}, \bibinfo {author} {\bibfnamefont {A.}~\bibnamefont
  {{Carlini}}}, \bibinfo {author} {\bibfnamefont {A.}~\bibnamefont {{Mari}}},
  \bibinfo {author} {\bibfnamefont {T.}~\bibnamefont {{Caneva}}}, \bibinfo
  {author} {\bibfnamefont {S.}~\bibnamefont {{Montangero}}}, \bibinfo {author}
  {\bibfnamefont {T.}~\bibnamefont {{Calarco}}}, \bibinfo {author}
  {\bibfnamefont {R.}~\bibnamefont {{Fazio}}}, \ and\ \bibinfo {author}
  {\bibfnamefont {V.}~\bibnamefont {{Giovannetti}}},\ }\href {\doibase
  10.1103/PhysRevA.88.062326} {\bibfield  {journal} {\bibinfo  {journal}
  {\PRA}\ }\textbf {\bibinfo {volume} {88}},\ \bibinfo {eid} {062326} (\bibinfo
  {year} {2013})}\BibitemShut {NoStop}%
\bibitem [{\citenamefont {{Xu}}\ \emph {et~al.}(2014)\citenamefont {{Xu}},
  \citenamefont {{Luo}}, \citenamefont {{Yang}}, \citenamefont {{Liu}},\ and\
  \citenamefont {{Zhu}}}]{Xu2014}%
  \BibitemOpen
  \bibfield  {author} {\bibinfo {author} {\bibfnamefont {Z.-Y.}\ \bibnamefont
  {{Xu}}}, \bibinfo {author} {\bibfnamefont {S.}~\bibnamefont {{Luo}}},
  \bibinfo {author} {\bibfnamefont {W.~L.}\ \bibnamefont {{Yang}}}, \bibinfo
  {author} {\bibfnamefont {C.}~\bibnamefont {{Liu}}}, \ and\ \bibinfo {author}
  {\bibfnamefont {S.}~\bibnamefont {{Zhu}}},\ }\href {\doibase
  10.1103/PhysRevA.89.012307} {\bibfield  {journal} {\bibinfo  {journal}
  {\PRA}\ }\textbf {\bibinfo {volume} {89}},\ \bibinfo {eid} {012307} (\bibinfo
  {year} {2014})}\BibitemShut {NoStop}%
\bibitem [{\citenamefont {{Xu}}\ and\ \citenamefont {{Zhu}}(2014)}]{Xu2014a}%
  \BibitemOpen
  \bibfield  {author} {\bibinfo {author} {\bibfnamefont {Z.-Y.}\ \bibnamefont
  {{Xu}}}\ and\ \bibinfo {author} {\bibfnamefont {S.-Q.}\ \bibnamefont
  {{Zhu}}},\ }\href {\doibase 10.1088/0256-307X/31/2/020301} {\bibfield
  {journal} {\bibinfo  {journal} {Chin. Phys. Lett.}\ }\textbf {\bibinfo
  {volume} {31}},\ \bibinfo {eid} {020301} (\bibinfo {year}
  {2014})}\BibitemShut {NoStop}%
%\bibitem [{\citenamefont {{Richerme}}\ \emph {et~al.}(2014)\citenamefont
%  {{Richerme}}, \citenamefont {{Gong}}, \citenamefont {{Lee}}, \citenamefont
%  {{Senko}}, \citenamefont {{Smith}}, \citenamefont {{Foss-Feig}},
%  \citenamefont {{Michalakis}}, \citenamefont {{Gorshkov}},\ and\ \citenamefont
%  {{Monroe}}}]{richerme14}%
%  \BibitemOpen
%  \bibfield  {author} {\bibinfo {author} {\bibfnamefont {P.}~\bibnamefont
%  {{Richerme}}}, \bibinfo {author} {\bibfnamefont {Z.-X.}\ \bibnamefont
%  {{Gong}}}, \bibinfo {author} {\bibfnamefont {A.}~\bibnamefont {{Lee}}},
%  \bibinfo {author} {\bibfnamefont {C.}~\bibnamefont {{Senko}}}, \bibinfo
%  {author} {\bibfnamefont {J.}~\bibnamefont {{Smith}}}, \bibinfo {author}
%  {\bibfnamefont {M.}~\bibnamefont {{Foss-Feig}}}, \bibinfo {author}
%  {\bibfnamefont {S.}~\bibnamefont {{Michalakis}}}, \bibinfo {author}
%  {\bibfnamefont {A.~V.}\ \bibnamefont {{Gorshkov}}}, \ and\ \bibinfo {author}
%  {\bibfnamefont {C.}~\bibnamefont {{Monroe}}},\ }\href
%  {http://www.nature.com/nature/journal/v511/n7508/full/nature13450.html}
%  {\bibfield  {journal} {\bibinfo  {journal} {Nature}\ }\textbf {\bibinfo
%  {volume} {511}},\ \bibinfo {pages} {198} (\bibinfo {year}
%  {2014})}\BibitemShut {NoStop}%
%\bibitem [{\citenamefont {Jurcevic}\ \emph {et~al.}(2014)\citenamefont
%  {Jurcevic}, \citenamefont {Lanyon}, \citenamefont {Hauke}, \citenamefont
%  {Hempel}, \citenamefont {Zoller}, \citenamefont {Blatt},\ and\ \citenamefont
%  {Roos}}]{jurcevic14}%
%  \BibitemOpen
%  \bibfield  {author} {\bibinfo {author} {\bibfnamefont {P.}~\bibnamefont
%  {Jurcevic}}, \bibinfo {author} {\bibfnamefont {B.~P.}\ \bibnamefont
%  {Lanyon}}, \bibinfo {author} {\bibfnamefont {P.}~\bibnamefont {Hauke}},
%  \bibinfo {author} {\bibfnamefont {C.}~\bibnamefont {Hempel}}, \bibinfo
%  {author} {\bibfnamefont {P.}~\bibnamefont {Zoller}}, \bibinfo {author}
%  {\bibfnamefont {R.}~\bibnamefont {Blatt}}, \ and\ \bibinfo {author}
%  {\bibfnamefont {C.~F.}\ \bibnamefont {Roos}},\ }\href
%  {http://dx.doi.org/10.1038/nature13461} {\bibfield  {journal} {\bibinfo
%  {journal} {Nature}\ }\textbf {\bibinfo {volume} {511}},\ \bibinfo {pages}
%  {202} (\bibinfo {year} {2014})}\BibitemShut {NoStop}%
\bibitem [{\citenamefont {Madsen}\ \emph {et~al.}(2011)\citenamefont {Madsen},
  \citenamefont {Ates}, \citenamefont {Lund-Hansen}, \citenamefont {L\"offler},
  \citenamefont {Reitzenstein}, \citenamefont {Forchel},\ and\ \citenamefont
  {Lodahl}}]{madsen11}%
  \BibitemOpen
  \bibfield  {author} {\bibinfo {author} {\bibfnamefont {K.~H.}\ \bibnamefont
  {Madsen}}, \bibinfo {author} {\bibfnamefont {S.}~\bibnamefont {Ates}},
  \bibinfo {author} {\bibfnamefont {T.}~\bibnamefont {Lund-Hansen}}, \bibinfo
  {author} {\bibfnamefont {A.}~\bibnamefont {L\"offler}}, \bibinfo {author}
  {\bibfnamefont {S.}~\bibnamefont {Reitzenstein}}, \bibinfo {author}
  {\bibfnamefont {A.}~\bibnamefont {Forchel}}, \ and\ \bibinfo {author}
  {\bibfnamefont {P.}~\bibnamefont {Lodahl}},\ }\href {\doibase
  10.1103/PhysRevLett.106.233601} {\bibfield  {journal} {\bibinfo  {journal}
  {Phys. Rev. Lett.}\ }\textbf {\bibinfo {volume} {106}},\ \bibinfo {pages}
  {233601} (\bibinfo {year} {2011})}\BibitemShut {NoStop}%
\bibitem [{\citenamefont {Carmichael}\ \emph {et~al.}(1991)\citenamefont
  {Carmichael}, \citenamefont {Brecha},\ and\ \citenamefont
  {Rice}}]{carmichael91}%
  \BibitemOpen
  \bibfield  {author} {\bibinfo {author} {\bibfnamefont {H.~J.}\ \bibnamefont
  {Carmichael}}, \bibinfo {author} {\bibfnamefont {R.~J.}\ \bibnamefont
  {Brecha}}, \ and\ \bibinfo {author} {\bibfnamefont {P.~R.}\ \bibnamefont
  {Rice}},\ }\href
  {http://www.sciencedirect.com/science/article/pii/003040189190194I}
  {\bibfield  {journal} {\bibinfo  {journal} {Opt. Commun.}\ }\textbf {\bibinfo
  {volume} {82}},\ \bibinfo {pages} {73} (\bibinfo {year} {1991})}\BibitemShut
  {NoStop}%
\bibitem [{\citenamefont {Rempe}\ \emph {et~al.}(1991)\citenamefont {Rempe},
  \citenamefont {Thompson}, \citenamefont {Brecha}, \citenamefont {Lee},\ and\
  \citenamefont {Kimble}}]{rempe91}%
  \BibitemOpen
  \bibfield  {author} {\bibinfo {author} {\bibfnamefont {G.}~\bibnamefont
  {Rempe}}, \bibinfo {author} {\bibfnamefont {R.~J.}\ \bibnamefont {Thompson}},
  \bibinfo {author} {\bibfnamefont {R.~J.}\ \bibnamefont {Brecha}}, \bibinfo
  {author} {\bibfnamefont {W.~D.}\ \bibnamefont {Lee}}, \ and\ \bibinfo
  {author} {\bibfnamefont {H.~J.}\ \bibnamefont {Kimble}},\ }\href {\doibase
  10.1103/PhysRevLett.67.1727} {\bibfield  {journal} {\bibinfo  {journal}
  {Phys. Rev. Lett.}\ }\textbf {\bibinfo {volume} {67}},\ \bibinfo {pages}
  {1727} (\bibinfo {year} {1991})}\BibitemShut {NoStop}%
\bibitem [{\citenamefont {Mielke}\ \emph {et~al.}(1998)\citenamefont {Mielke},
  \citenamefont {Foster},\ and\ \citenamefont {Orozco}}]{mielke98}%
  \BibitemOpen
  \bibfield  {author} {\bibinfo {author} {\bibfnamefont {S.~L.}\ \bibnamefont
  {Mielke}}, \bibinfo {author} {\bibfnamefont {G.~T.}\ \bibnamefont {Foster}},
  \ and\ \bibinfo {author} {\bibfnamefont {L.~A.}\ \bibnamefont {Orozco}},\
  }\href {http://journals.aps.org/prl/abstract/10.1103/PhysRevLett.80.3948}
  {\bibfield  {journal} {\bibinfo  {journal} {Phys. Rev. Lett.}\ }\textbf
  {\bibinfo {volume} {80}},\ \bibinfo {pages} {3948} (\bibinfo {year}
  {1998})}\BibitemShut {NoStop}%
\bibitem [{\citenamefont {Foster}\ \emph
  {et~al.}(2000{\natexlab{a}})\citenamefont {Foster}, \citenamefont {Mielke},\
  and\ \citenamefont {Orozco}}]{foster00pra}%
  \BibitemOpen
  \bibfield  {author} {\bibinfo {author} {\bibfnamefont {G.~T.}\ \bibnamefont
  {Foster}}, \bibinfo {author} {\bibfnamefont {S.~L.}\ \bibnamefont {Mielke}},
  \ and\ \bibinfo {author} {\bibfnamefont {L.~A.}\ \bibnamefont {Orozco}},\
  }\href {http://journals.aps.org/pra/abstract/10.1103/PhysRevA.61.053821}
  {\bibfield  {journal} {\bibinfo  {journal} {Phys. Rev. A}\ }\textbf {\bibinfo
  {volume} {61}},\ \bibinfo {pages} {053821} (\bibinfo {year}
  {2000}{\natexlab{a}})}\BibitemShut {NoStop}%
\bibitem [{\citenamefont {Thompson}\ \emph {et~al.}(1992)\citenamefont
  {Thompson}, \citenamefont {Rempe},\ and\ \citenamefont
  {Kimble}}]{thompson92}%
  \BibitemOpen
  \bibfield  {author} {\bibinfo {author} {\bibfnamefont {R.~J.}\ \bibnamefont
  {Thompson}}, \bibinfo {author} {\bibfnamefont {G.}~\bibnamefont {Rempe}}, \
  and\ \bibinfo {author} {\bibfnamefont {H.~J.}\ \bibnamefont {Kimble}},\
  }\href {http://journals.aps.org/prl/abstract/10.1103/PhysRevLett.68.1132}
  {\bibfield  {journal} {\bibinfo  {journal} {Phys. Rev. Lett.}\ }\textbf
  {\bibinfo {volume} {68}},\ \bibinfo {pages} {1132} (\bibinfo {year}
  {1992})}\BibitemShut {NoStop}%
\bibitem [{\citenamefont {Carmichael}\ and\ \citenamefont
  {Sanders}(1999)}]{carmichael99}%
  \BibitemOpen
  \bibfield  {author} {\bibinfo {author} {\bibfnamefont {H.~J.}\ \bibnamefont
  {Carmichael}}\ and\ \bibinfo {author} {\bibfnamefont {B.~C.}\ \bibnamefont
  {Sanders}},\ }\href
  {http://journals.aps.org/pra/abstract/10.1103/PhysRevA.60.2497} {\bibfield
  {journal} {\bibinfo  {journal} {Phys. Rev. A}\ }\textbf {\bibinfo {volume}
  {60}},\ \bibinfo {pages} {2497} (\bibinfo {year} {1999})}\BibitemShut
  {NoStop}%
%\bibitem [{\citenamefont {Reiner}\ \emph {et~al.}(2004)\citenamefont {Reiner},
%  \citenamefont {Dimler},\ and\ \citenamefont {Orozco}}]{reiner04b}%
%  \BibitemOpen
%  \bibfield  {author} {\bibinfo {author} {\bibfnamefont {J.~E.}\ \bibnamefont
%  {Reiner}}, \bibinfo {author} {\bibfnamefont {F.~M.}\ \bibnamefont {Dimler}},
%  \ and\ \bibinfo {author} {\bibfnamefont {L.~A.}\ \bibnamefont {Orozco}},\
%  }\href {http://iopscience.iop.org/1464-4266/6/2/003} {\bibfield  {journal}
%  {\bibinfo  {journal} {J. Opt. B: Quantum Semiclass. Opt.}\ }\textbf {\bibinfo
%  {volume} {6}},\ \bibinfo {pages} {135} (\bibinfo {year} {2004})}\BibitemShut
%  {NoStop}%
\bibitem{brecha99} R. J. Brecha,  P. R. Rice, P. R. and M. Xiao, Phys. Rev. A {\bf 59}, 2392 (1999).
\bibitem [{\citenamefont {Norris}\ \emph {et~al.}(2009)\citenamefont {Norris},
  \citenamefont {Cahoon},\ and\ \citenamefont {Orozco}}]{norris09a}%
  \BibitemOpen
  \bibfield  {author} {\bibinfo {author} {\bibfnamefont {D.~G.}\ \bibnamefont
  {Norris}}, \bibinfo {author} {\bibfnamefont {E.~J.}\ \bibnamefont {Cahoon}},
  \ and\ \bibinfo {author} {\bibfnamefont {L.~A.}\ \bibnamefont {Orozco}},\
  }\href {http://journals.aps.org/pra/abstract/10.1103/PhysRevA.80.043830}
  {\bibfield  {journal} {\bibinfo  {journal} {Phys. Rev. A}\ }\textbf {\bibinfo
  {volume} {80}},\ \bibinfo {pages} {{043830}} (\bibinfo {year}
  {2009})}\BibitemShut {NoStop}%
\bibitem [{\citenamefont {Kulin}\ \emph {et~al.}(2001)\citenamefont {Kulin},
  \citenamefont {Aubin}, \citenamefont {Christe}, \citenamefont {Peker},
  \citenamefont {Rolston},\ and\ \citenamefont {Orozco}}]{kulin01}%
  \BibitemOpen
  \bibfield  {author} {\bibinfo {author} {\bibfnamefont {S.}~\bibnamefont
  {Kulin}}, \bibinfo {author} {\bibfnamefont {S.}~\bibnamefont {Aubin}},
  \bibinfo {author} {\bibfnamefont {S.}~\bibnamefont {Christe}}, \bibinfo
  {author} {\bibfnamefont {B.}~\bibnamefont {Peker}}, \bibinfo {author}
  {\bibfnamefont {S.~L.}\ \bibnamefont {Rolston}}, \ and\ \bibinfo {author}
  {\bibfnamefont {L.~A.}\ \bibnamefont {Orozco}},\ }\href
  {http://stacks.iop.org/1464-4266/3/i=6/a=301} {\bibfield  {journal} {\bibinfo
   {journal} {J. Opt. B: Quantum Semiclass. Opt.}\ }\textbf {\bibinfo {volume}
  {3}},\ \bibinfo {pages} {353} (\bibinfo {year} {2001})}\BibitemShut {NoStop}%
\bibitem [{\citenamefont {Zhu}\ \emph {et~al.}(1990)\citenamefont {Zhu},
  \citenamefont {Gauthier}, \citenamefont {Morin}, \citenamefont {Wu},
  \citenamefont {Carmichael},\ and\ \citenamefont {Mossberg}}]{zhu90}%
  \BibitemOpen
  \bibfield  {author} {\bibinfo {author} {\bibfnamefont {Y.}~\bibnamefont
  {Zhu}}, \bibinfo {author} {\bibfnamefont {D.~J.}\ \bibnamefont {Gauthier}},
  \bibinfo {author} {\bibfnamefont {S.~E.}\ \bibnamefont {Morin}}, \bibinfo
  {author} {\bibfnamefont {Q.}~\bibnamefont {Wu}}, \bibinfo {author}
  {\bibfnamefont {H.~J.}\ \bibnamefont {Carmichael}}, \ and\ \bibinfo {author}
  {\bibfnamefont {T.~W.}\ \bibnamefont {Mossberg}},\ }\href
  {http://journals.aps.org/prl/abstract/10.1103/PhysRevLett.64.2499} {\bibfield
   {journal} {\bibinfo  {journal} {Phys. Rev. Lett.}\ }\textbf {\bibinfo
  {volume} {64}},\ \bibinfo {pages} {2499} (\bibinfo {year}
  {1990})}\BibitemShut {NoStop}%
\bibitem [{\citenamefont {Gripp}\ \emph {et~al.}(1997)\citenamefont {Gripp},
  \citenamefont {Mielke},\ and\ \citenamefont {Orozco}}]{gripp97}%
  \BibitemOpen
  \bibfield  {author} {\bibinfo {author} {\bibfnamefont {J.}~\bibnamefont
  {Gripp}}, \bibinfo {author} {\bibfnamefont {S.~L.}\ \bibnamefont {Mielke}}, \
  and\ \bibinfo {author} {\bibfnamefont {L.~A.}\ \bibnamefont {Orozco}},\
  }\href {http://journals.aps.org/pra/abstract/10.1103/PhysRevA.56.3262}
  {\bibfield  {journal} {\bibinfo  {journal} {Phys. Rev. A}\ }\textbf {\bibinfo
  {volume} {56}},\ \bibinfo {pages} {3262} (\bibinfo {year}
  {1997})}\BibitemShut {NoStop}%
%\bibitem [{\citenamefont {Foster}\ \emph
%  {et~al.}(2000{\natexlab{b}})\citenamefont {Foster}, \citenamefont {Orozco},
%  \citenamefont {Castro-Beltran},\ and\ \citenamefont {Carmichael}}]{foster00}%
%  \BibitemOpen
%  \bibfield  {author} {\bibinfo {author} {\bibfnamefont {G.~T.}\ \bibnamefont
%  {Foster}}, \bibinfo {author} {\bibfnamefont {L.~A.}\ \bibnamefont {Orozco}},
%  \bibinfo {author} {\bibfnamefont {H.~M.}\ \bibnamefont {Castro-Beltran}}, \
%  and\ \bibinfo {author} {\bibfnamefont {H.~J.}\ \bibnamefont {Carmichael}},\
%  }\href {http://prl.aps.org/abstract/PRL/v85/i15/p3149_1} {\bibfield
%  {journal} {\bibinfo  {journal} {Phys. Rev. Lett.}\ }\textbf {\bibinfo
%  {volume} {85}},\ \bibinfo {pages} {3149} (\bibinfo {year}
%  {2000}{\natexlab{b}})}\BibitemShut {NoStop}%
%\bibitem [{Note1()}]{Note1}%
%  \BibitemOpen
%  \bibinfo {note} {These results together with findings for multi-level systems and {\color{red} a model including the effects of the velocity distribute of the atoms} will be published elsewhere}\BibitemShut {NoStop}%
\bibitem [{\citenamefont {Brune}\ \emph {et~al.}(1996)\citenamefont {Brune},
  \citenamefont {Schmidt-Kaler}, \citenamefont {Maali}, \citenamefont {Dreyer},
  \citenamefont {Hagley}, \citenamefont {\mbox{Raimond}},\ and\ \citenamefont
  {Haroche}}]{brune96}%
  \BibitemOpen
  \bibfield  {author} {\bibinfo {author} {\bibfnamefont {M.}~\bibnamefont
  {Brune}}, \bibinfo {author} {\bibfnamefont {F.}~\bibnamefont
  {Schmidt-Kaler}}, \bibinfo {author} {\bibfnamefont {A.}~\bibnamefont
  {Maali}}, \bibinfo {author} {\bibfnamefont {J.}~\bibnamefont {Dreyer}},
  \bibinfo {author} {\bibfnamefont {E.}~\bibnamefont {Hagley}}, \bibinfo
  {author} {\bibfnamefont {J.~M.}\ \bibnamefont {\mbox{Raimond}}}, \ and\
  \bibinfo {author} {\bibfnamefont {S.}~\bibnamefont {Haroche}},\ }\href
  {http://journals.aps.org/prl/abstract/10.1103/PhysRevLett.76.1800} {\bibfield
   {journal} {\bibinfo  {journal} {Phys. Rev. Lett.}\ }\textbf {\bibinfo
  {volume} {76}},\ \bibinfo {pages} {1800} (\bibinfo {year}
  {1996})}\BibitemShut {NoStop}%
\bibitem [{\citenamefont {Breuer}\ and\ \citenamefont
  {Petruccione}(2007)}]{breuer07book}%
  \BibitemOpen
  \bibfield  {author} {\bibinfo {author} {\bibfnamefont {H.-P.}\ \bibnamefont
  {Breuer}}\ and\ \bibinfo {author} {\bibfnamefont {F.}~\bibnamefont
  {Petruccione}},\ }\href@noop {} {\emph {\bibinfo {title} {The Theory of Open
  Quantum Systems}}}\ (\bibinfo  {publisher} {Oxford University Press},\
  \bibinfo {address} {Oxford},\ \bibinfo {year} {2007})\BibitemShut {NoStop}%
\bibitem [{\citenamefont {Breuer}\ \emph {et~al.}(2009)\citenamefont {Breuer},
  \citenamefont {Laine},\ and\ \citenamefont {Piilo}}]{breuer09}%
  \BibitemOpen
  \bibfield  {author} {\bibinfo {author} {\bibfnamefont {H.-P.}\ \bibnamefont
  {Breuer}}, \bibinfo {author} {\bibfnamefont {E.-M.}\ \bibnamefont {Laine}}, \
  and\ \bibinfo {author} {\bibfnamefont {J.}~\bibnamefont {Piilo}},\ }\href
  {\doibase 10.1103/PhysRevLett.103.210401} {\bibfield  {journal} {\bibinfo
  {journal} {Phys. Rev. Lett.}\ }\textbf {\bibinfo {volume} {103}},\ \bibinfo
  {pages} {210401} (\bibinfo {year} {2009})}\BibitemShut {NoStop}%
\bibitem {Note2} It is worth emphasizing that in our study increase of
  non-Markovianity is a necessary, but not sufficient condition to observe a
  speed-up. See also Ref.~\cite {Xu2014}.
%  \bibitem [{Note2()}]{Note2}%
%  \BibitemOpen
%  \bibinfo {note} {It is worth emphasizing that in our study increase of
%  non-Markovianity is a necessary, but not sufficient condition to observe a
%  speed-up. See also Ref.~\cite {Xu2014}}\BibitemShut {NoStop}%
\bibitem [{\citenamefont {Smith}\ \emph {et~al.}(2002)\citenamefont {Smith},
  \citenamefont {Reiner}, \citenamefont {Orozco}, \citenamefont {Kuhr},\ and\
  \citenamefont {Wiseman}}]{smith02}%
  \BibitemOpen
  \bibfield  {author} {\bibinfo {author} {\bibfnamefont {W.~P.}\ \bibnamefont
  {Smith}}, \bibinfo {author} {\bibfnamefont {J.~E.}\ \bibnamefont {Reiner}},
  \bibinfo {author} {\bibfnamefont {L.~A.}\ \bibnamefont {Orozco}}, \bibinfo
  {author} {\bibfnamefont {S.}~\bibnamefont {Kuhr}}, \ and\ \bibinfo {author}
  {\bibfnamefont {H.~M.}\ \bibnamefont {Wiseman}},\ }\href
  {http://journals.aps.org/prl/abstract/10.1103/PhysRevLett.89.133601}
  {\bibfield  {journal} {\bibinfo  {journal} {Phys. Rev. Lett.}\ }\textbf
  {\bibinfo {volume} {89}},\ \bibinfo {pages} {{133601}} (\bibinfo {year}
  {2002})}\BibitemShut {NoStop}%
\bibitem [{\citenamefont {Cimmarusti}\ \emph {et~al.}(2013)\citenamefont
  {Cimmarusti}, \citenamefont {Schroeder}, \citenamefont {Patterson},
  \citenamefont {Orozco}, \citenamefont {Barberis-Blostein},\ and\
  \citenamefont {Carmichael}}]{cimmarusti13}%
  \BibitemOpen
  \bibfield  {author} {\bibinfo {author} {\bibfnamefont {A.~D.}\ \bibnamefont
  {Cimmarusti}}, \bibinfo {author} {\bibfnamefont {C.~A.}\ \bibnamefont
  {Schroeder}}, \bibinfo {author} {\bibfnamefont {B.~D.}\ \bibnamefont
  {Patterson}}, \bibinfo {author} {\bibfnamefont {L.~A.}\ \bibnamefont
  {Orozco}}, \bibinfo {author} {\bibfnamefont {P.}~\bibnamefont
  {Barberis-Blostein}}, \ and\ \bibinfo {author} {\bibfnamefont {H.~J.}\
  \bibnamefont {Carmichael}},\ }\href
  {http://stacks.iop.org/1367-2630/15/i=1/a=013017} {\bibfield  {journal}
  {\bibinfo  {journal} {New J. Phys.}\ }\textbf {\bibinfo {volume} {15}},\
  \bibinfo {pages} {{013017}} (\bibinfo {year} {2013})}\BibitemShut {NoStop}%
%\bibitem [{\citenamefont {Norris}\ \emph {et~al.}(2010)\citenamefont {Norris},
%  \citenamefont {Orozco}, \citenamefont {Barberis-Blostein},\ and\
%  \citenamefont {Carmichael}}]{norris10}%
%  \BibitemOpen
%  \bibfield  {author} {\bibinfo {author} {\bibfnamefont {D.~G.}\ \bibnamefont
%  {Norris}}, \bibinfo {author} {\bibfnamefont {L.~A.}\ \bibnamefont {Orozco}},
%  \bibinfo {author} {\bibfnamefont {P.}~\bibnamefont {Barberis-Blostein}}, \
%  and\ \bibinfo {author} {\bibfnamefont {H.~J.}\ \bibnamefont {Carmichael}},\
%  }\href {\doibase 10.1103/PhysRevLett.105.123602} {\bibfield  {journal}
%  {\bibinfo  {journal} {Phys. Rev. Lett.}\ }\textbf {\bibinfo {volume}
%  {{105}}},\ \bibinfo {pages} {{123602}} (\bibinfo {year} {2010})}\BibitemShut
%  {NoStop}%
%\bibitem [{\citenamefont {Norris}\ \emph {et~al.}(2012)\citenamefont {Norris},
%  \citenamefont {Cimmarusti}, \citenamefont {Orozco}, \citenamefont
%  {Barberis-Blostein},\ and\ \citenamefont {Carmichael}}]{norris12b}%
%  \BibitemOpen
%  \bibfield  {author} {\bibinfo {author} {\bibfnamefont {D.~G.}\ \bibnamefont
%  {Norris}}, \bibinfo {author} {\bibfnamefont {A.~D.}\ \bibnamefont
%  {Cimmarusti}}, \bibinfo {author} {\bibfnamefont {L.~A.}\ \bibnamefont
%  {Orozco}}, \bibinfo {author} {\bibfnamefont {P.}~\bibnamefont
%  {Barberis-Blostein}}, \ and\ \bibinfo {author} {\bibfnamefont {H.~J.}\
%  \bibnamefont {Carmichael}},\ }\href {\doibase 10.1103/PhysRevA.86.053816}
%  {\bibfield  {journal} {\bibinfo  {journal} {Phys. Rev. A}\ }\textbf {\bibinfo
%  {volume} {86}},\ \bibinfo {pages} {{053816}} (\bibinfo {year}
%  {2012})}\BibitemShut {NoStop}%
\end{thebibliography}
\end{document}